\documentclass[aoas,MSNbibl,nameyear,seceqn,dvips]{arximspdf}
\usepackage{dcolumn}
\usepackage{graphicx}

\doi{10.1214/14-AOAS752} 
\volume{8}
\issue{3}
\pubyear{2014}
\firstpage{1713}
\lastpage{1727}
\docsubty{FLA}

\makeatletter
\newcolumntype{d}[1]{D{.}{.}{#1}}
\newcommand{\bmu}{\bolds{\mu}}
\newcommand{\bSigma}{\bolds{\Sigma}}
\makeatother

\begin{document}
\begin{frontmatter}

\title{On analysis of incomplete field failure data}
\runtitle{Incomplete field failure data}

\begin{aug}
\author[A]{\fnms{Zhisheng}~\snm{Ye}\ead[label=e1]{yez@nus.edu.sg}\thanksref{T1}}
\and
\author[B]{\fnms{Hon Keung Tony}~\snm{Ng}\corref{}\ead[label=e2]{ngh@mail.smu.edu}\thanksref{T2}}
\runauthor{Z. Ye and H. K. T. Ng}
\affiliation{National University of Singapore and Southern Methodist University}
\address[A]{Department of Industrial\\
\quad and System Engineering\\
National University of Singapore\\
1 Engineering Drive 2\\
Singapore 117576\\
Republic of Singapore\\
\printead{e1}}
\address[B]{Department of Statistical Science\\
Southern Methodist University\\
Dallas, Texas 75275-0332\\
USA\\
\printead{e2}}
\end{aug}
\thankstext{T1}{Supported by the National Research Foundation
Singapore under its Campus for Research Excellence and Technological
Enterprise (CREATE).}
\thankstext{T2}{Supported by a Grant from the Simons Foundation (\#
280601 to Tony Ng).}

\received{\smonth{11} \syear{2013}}
\revised{\smonth{5} \syear{2014}}

%
\begin{abstract}
Many commercial products are sold with warranties and indirectly
through dealers. The manufacturer-retailer distribution mechanism
results in serious missing data problems in field return data, as the
sales date for an unreturned unit is generally unknown to the
manufacturer. This study considers a general setting for field failure
data with unknown sales dates and a warranty limit. A stochastic
expectation--maximization (SEM) algorithm is developed to estimate the
distributions of the sales lag (time between shipment to a retailer and
sale to a customer) and the lifetime of the product under study.
Extensive simulations are used to evaluate the performance of the SEM
algorithm and to compare with the imputation method proposed by Ghosh
[\textit{Ann. Appl. Stat.} \textbf{4} (2010) 1976--1999].
Three real examples illustrate the methodology proposed in this paper.
\end{abstract}

%
\end{frontmatter}

\setcounter{footnote}{2}

\section{Introduction}\label{sec1}

Field failure data contain rich information about product
reliability and the operating conditions in actual use. The information
is important for risk assessment of field failures, early detection of
unanticipated reliability problems [\citet{482}], and prediction
of operation costs. Since many commercial products are sold with
warranties, field failure data usually come from warranty claims.
Alternatively, for noncommercial products such as military products,
field data may be extracted from maintenance reports [\citet
{346}], and this type of data is called field maintenance data.

The rich information contained in field failure data can be extracted
by careful data analysis. However, the analysis is difficult because
field data are generally coarse and of poor quality. Compared with lab
data that are collected under well-controlled testing conditions, field
data are collected from customers and are contaminated by customer
behaviors. For instance, the data are often contaminated with
heterogeneous use conditions [\citet{ye2013}], dormant period
after purchase [\citet{1012}], delayed report after failure
[\citet{538}], customer rush near warranty expiration
[\citet{601}], the failed-but-not-reported problem [\citet
{xie13}], and systematic error on the cause and time of failures due to
report error. To address these issues, a number of statistical models
have been developed for warranty data analysis. See \citet{1015}
and \citet{1012} for a comprehensive overview.

Another important cause of the coarse data is the missing sales date of
unreturned units. Nowadays, many products are sold to customers through
multiple channels of distribution instead of direct sale from the
manufacturers. Under the manufacturer-retailer distribution mechanism,
if a product fails within warranty, it will be returned to the
manufacturer as a warranty claim. Then, the lifetime and the sales lag,
which is the time between shipment to a retailer and sale to a
customer, can be easily obtained from the warranty card. For an
unreturned unit, however, the sales date is generally unknown unless
the product is expensive (e.g., cars). The unit might still be in a
retailer's warehouse or it might have been sold to a customer at some
date unknown to the manufacturer.
\citet{1014} presented such an example, where residential furnace
components were shipped to retailers in batches and then sold to
customers through retailers. Because of the retailers, the exact
release time of a furnace to a customer was generally masked unless a
furnace was sold and failed before a fixed end-of-study date.

A common approach to the unknown sales lag problem is to carry out a
sensitivity analysis by assuming that the sales lag is fixed
[\citet{lawless98}]. Another method is to obtain the sales-lag
distribution using survey or past experience, and then this
distribution is incorporated into the data analysis to improve
estimation accuracy [\citet{542,1011}]. This method does not make
full use of the database, as the sales date for returned units can be
read from the warranty card and the sales-lag information is available
from these returned units. Some studies treat both the observed
sales-lag data and the observed lifetime as right censored so that the
two types of data can be analyzed separately [\citet
{ion07,karim2008,akbarov12}]. Given that the sales date of an
unreturned unit is unknown, however, the sales-lag data are not right
censored, and the lifetime data are neither left truncated nor right
censored. To get an accurate estimate, the sales-lag data and the
lifetime data have to be jointly analyzed. In an interesting study,
\citet{1014} analyzed field failure data with unknown sales lags.
However, the inference procedure in that work is not efficient. In
addition, it does not allow for a warranty limit and, thus, it is not
applicable to warranty data. In addition, previous\vadjust{\goodbreak} research assumes
independent sales lag and lifetime. This assumption is true for some
products, for example, light bulbs, televisions, computers, etc. For
seasonal products such as heaters, fans, and air purifiers, the sales
lag and the lifetime are correlated due to the usage pattern. For
instance, a heater sold in summer will last longer than one sold in
winter due to the uneven usage. It is also possible that a longer sales
lag introduces more damage to the product [\citet{akbarov12}].
Moreover, most research on field data analysis emphasizes the field
failure time distribution only. The sales-lag information reflects
customer demand rate and is important in manufacturing and inventory decisions.

In this paper, we consider joint parametric inference of sales-lag and
lifetime in the presence of unknown sales dates. In contrast to the
work by \citet{1014}, we allow for a warranty limit as well as
dependency between sales lag and lifetime. In addition, we propose a
more efficient algorithm for statistical inference. Section~\ref{sec2} presents
a simplified problem setting for field data with a warranty limit and
an end-of-study date. Section~\ref{sec3} proposes an inference framework based
on the stochastic expectation--maximization (SEM) algorithm. Section~\ref{sec4}
discusses how the SEM algorithm can be modified to handle more general
situations. In Section~\ref{sec5} a simulation study examines the performance of
the proposed algorithm, and we compare it with the imputation method
proposed by \citet{1014}. The proposed algorithm is demonstrated
using three examples with different missing data patterns in Section~\ref{sec6}.
A concise conclusion is provided in Section~\ref{sec7}.

\section{Problem statement}\label{sec2}\label{secproblem}
Suppose that $N$ identical units are produced in a batch and delivered
to several retailers at the same time. The delivery time is set as the
time origin in the analysis. These units are then sold to customers
with a warranty of length $\tau$, starting from the date of purchase.
Let $X$ be the sales lag (same as the sales date in this setting) and
$T$ the lifetime of the product from the date of sale, where both $X$
and $T$ are random.
Let $\mathcal T_0$ be a fixed end-of-study date, which can be viewed as
the date the analysis is performed. If a unit fails before $\mathcal
T_0$ and is within warranty, we assume that a warranty claim is made to
the manufacturer without delay. Then both the sales date $X$ and the
lifetime $T$ are known to us. Otherwise, the sales date and the product
lifetime are unavailable. Suppose that before $\mathcal T_0$, we
observe $C$ claims, and so we have $C$ realizations of $(X, T)$,
denoted as $(x_i, t_i)$, $i = 1, 2, \ldots, C$. For the remaining
$N-C$ units, the values of $(X, T)$ are missing.

This study focuses on parametric inference. Denote the joint
probability density function (PDF) of $(X, T)$ as $f_{X, T}(x, t)$ and
the joint cumulative distribution function (CDF) as $F_{X, T}(x, t)$,
where $x, t > 0$. Let $\Theta$ be the parameter vector. Given the
observed data $(x_i, t_i)$, $i = 1, 2, \ldots, C$, the likelihood
function of $\Theta$ is given by
%
\begin{equation}
\label{eqnlikelihood} \mathcal L (\Theta) = \bigl[1-\operatorname{Pr}(X + T < \mathcal
T_0, T < \tau) \bigr]^{N-C}\prod
_{i=1}^C{f_{X, T}(x_i,
t_i)},
\end{equation}
where $\operatorname{Pr}(X + T < \mathcal T_0, T < \tau)$ is the
probability that a unit fails within warranty and is observed within
$\mathcal T_0$. This probability can be written as
\[
\operatorname{Pr}(X + T < \mathcal T_0, T < \tau) = \int
_0^{\tau}\!{\int_0^{\mathcal T_0 - t}{f_{X, T}(x,
t)\,dx}\,dt}.
\]
If $X$ and $T$ are independent, this probability simplifies to
\[
\operatorname{Pr}(X + T < \mathcal T_0, T < \tau)=\int
_0^{\tau
}{F_X(\mathcal T_0
- t)\,dF_T(t)}.
\]
In principle, the maximum likelihood estimator (MLE) of $\Theta$ can
be obtained from direct maximization of the likelihood (\ref
{eqnlikelihood}). Nevertheless, numerical evaluation of the integral
would introduce computation error, which is magnified by the factor
$N-C$ in (\ref{eqnlikelihood}). Due to the high missing data rate in
our problem (i.e., large $N$ and small $C$), the total computation
error is significant, and the likelihood is flat near the maximum.
These two factors lead to unstable
estimates if direct maximization is used (i.e., convergence to values
far from the optimal or failure to converge). The instability is
observed in our simulation study (see Section~\ref{secsimulation})
and \citet{1014}. Therefore, alternative techniques are needed.
In the next section we propose an efficient and easy-to-implement
procedure based on the SEM algorithm.

\section{The stochastic expectation--maximization framework}\label{sec3}\label{secSEM}
%
\subsection{The SEM algorithm}\label{sec3.1}\label{secSEMintroduction}

The EM algorithm is an iterative procedure that repeatedly fills the
missing data in the complete-data log-likelihood with their conditional
expected values (E-step) and maximizes the complete data log-likelihood
to update the parameter estimates (M-step). The EM algorithm is
efficient in finding the MLEs when computation of the expectation and
the maximization are easy to perform. See \citet{773} for a
book-length account.
Unfortunately, the E-step is intractable when the EM algorithm is
applied to the problem in Section~\ref{secproblem}. Alternatively,
the expectation can be approximated through Monte Carlo simulation,
leading to the Monte Carlo EM (MCEM) algorithm. In our problem, the
approximation error of the expectation leads to a breakdown of the MCEM
algorithm because the likelihood is flat near the maximum.

The difficulty in executing the E-step can be efficiently addressed by
the SEM algorithm proposed by \citet{991}. The SEM algorithm
replaces the E-step with a stochastic step (S-step), which is easy to
implement as long as the missing data are easy to impute. Compared with
the MCEM algorithm, the SEM algorithm completes the observed sample by
replacing each missing datum with a value randomly drawn from the
distribution conditional on results from the previous step. The SEM
algorithm has been shown to be computationally less burdensome than the
MCEM algorithm. Because of the stochastic nature, it is free of the
saddle point problem, a serious problem for the EM algorithm
[\citet{996,994}].
It was shown by \citet{999}, \citet{1018} and \citet
{998} that  under suitable regularity
conditions the SEM estimators are efficient in the sense that the
variance approaches the Cram{\'e}r--Rao lower bound. Some applications
of the algorithm suggest that it is insensitive to starting values and
performs well for small or moderate sample sizes. See, for example,
\citet{1018}, \citet{994}, and \citet{1000}.


\subsection{Implementation}\label{sec3.2}\label{secSEMimplementation}

Let $\bolds\Omega$ and $\bolds\Gamma$ be the sets of observed and
missing data, respectively. {Here, $\Omega$ includes the $C$ observed
values of ($x_{i}, t_{i}$) and the information that $N-C$ observations
are missing.} Given the parameter values $\Theta^{(k)}$ of $\Theta$
from the $k$th SEM cycle, the $(k+1)$st cycle for the problem described
in Section~\ref{secproblem} evolves as follows:

\begin{longlist}[S-step.]
\item[\textit{S-step}.] Draw a random sample $\bolds\Gamma^{(k)}=\{
(x_j^{(k)},y_j^{(k)}); j=1, 2, \ldots, N-C\}$ from the conditional
distribution of $\{\bolds\Gamma|\bolds\Omega, \Theta^{(k)}\}$ to
update the pseudo $Q$-function
%
%
\begin{equation}
\label{eqnQfunction} Q(\Theta; \bolds\Omega,\bolds\Gamma)=\sum
_{i=1}^C{\ln f_{X,
T}(x_i,
t_i)}+ \sum_{j=1}^{N-C}{\ln
f_{X, T}\bigl(x_j^{(k)}, t_j^{(k)}
\bigr)}.
\end{equation}
\item[\textit{M-step}.] Maximize\vspace*{1pt} the pseudo $Q$-function (\ref
{eqnQfunction}), which is a complete data log-likelihood, to obtain
$\Theta^{(k+1)}$ for the next cycle.
\end{longlist}
The M-step deals with a complete-data log-likelihood. It is easy to
implement through direct optimization or with the help of statistical
software if some common distributions are used for $X$ and $T$, for
example, independent exponential, Weibull, or bivariate lognormal.
Under  suitable regularity conditions,
the sequence $\Theta^{(k)}$ converges to a random variable whose mean
is an asymptotically efficient estimator of $\Theta$.
These conditions typically are satisfied if the
complete data model and the missing data model are sufficiently smooth
[\citet{998}, Section~2.3]. The simulation results in
Section~\ref{secsimulation} support this argument for commonly used
lifetime distributions.
To obtain an estimate of $\Theta$, we run the SEM algorithm to obtain
$\Theta^{(k)}, k = 1, 2, \ldots, K$, discard the first few iterations
for burn-in, and average over the estimates from the remaining
iterations to get $\hat\Theta$. According to some reports [e.g.,
\citet{1019}] as well as our experience, a burn-in period of 100
cycles is long enough under moderate missing data rates, while an
additional 1000 iterations are sufficient to estimate $\Theta$.
Nevertheless, we suggest a trace plot of the $\{\Theta^{(k)}\}$
sequence versus the iterations for checking the sufficiency of the
burn-in, and determining a more appropriate burn-in duration, if necessary.

There are several ways to impute the missing data in the S-step. The
standard method is based on the conditional distribution of the
unobserved $(X,T)$, which is
%
%
\begin{equation}
\label{eqnPDFmissingData}
\qquad g_{X, T}(x, t)=\frac{f_{X, T}(x, t)}{1-\operatorname{Pr}(X + T <
\mathcal T_0, T < \tau)}\bigl(1-
  I\{ x+t<\mathcal T_0, t<\tau\}
\bigr),
\end{equation}
%
where $I\{\cdot\}$ is the indicator function. 
Direct sampling from this conditional PDF is difficult. We might resort
to the Markov chain Monte Carlo (MCMC) method. However, it is
inefficient to imbed an iterative algorithm (MCMC) into another one
(SEM). Due to the extremely high missing data rate in our problem, we
impute missing data in a natural way, which is somewhat brute force,
yet very straightforward, easy to implement, and efficient.

Recall that a unit is observed only when $X + T < \mathcal T_0$ and $T
< \tau$, while the probability of being observed is typically low.
This motivates us to impute the missing data $\bolds\Gamma^{(k)}$ by
using a simple acceptance-rejection method: an imputation $(x,t)$ from
$f_{X, T} ( x, t | \Theta^{(k)} ) $ is
rejected only when $x + t < \mathcal T_0$ and $t < \tau$. It can be
easily shown that $(X, T)$ imputed from this sampling scheme follows
the distribution given in (\ref{eqnPDFmissingData}). To use this
imputation scheme, a starting point $\Theta^{(0)}$ that leads to a
large mean value of $X$ or $T$ is strongly recommended in order to
avoid a high rejection rate at the outset of the SEM algorithm.
According to our comprehensive simulation trials, this scheme is very
efficient because the missing data rate, which approximately equals 1
minus the rejection rate, is high in our setting. The rejection rate
should be low as long as $\Theta^{(k)}$ is not too far away from the
true value.
 Therefore, the brute-force imputation
is expected to be effective in the sense that the computational time
for each SEM iteration is relatively small.

\subsection{Confidence intervals}\label{sec3.3}\label{secCIconstruction}

The log-likelihood based on full data $\mathbf D = \bolds\Omega\cup
\bolds\Gamma$ is the same as (\ref{eqnQfunction}). Because of the
simple structure of the full data likelihood, the score function and
the observed information matrix based on full data can be easily
obtained by taking the first and second derivatives of (\ref
{eqnQfunction}) with respect to the parameters $\Theta$. Denote the
first and the negative of the second derivatives as $S(\Theta,\mathbf
D)$ and $B(\Theta,\mathbf D)$, respectively.
The observed information matrix based on incomplete data can be
computed based on the missing information principle [\citet
{louis82}] as
%
%
\begin{equation}
\label{eqnincompleteDataOIM} \mathcal I(\Theta) = E\bigl[B(\Theta,\mathbf D)|\bolds\Omega\bigr]
- E\bigl[S^2(\Theta,\mathbf D)|\bolds\Omega\bigr] + \bigl\{E\bigl[S(
\Theta,\mathbf D)|\bolds\Omega\bigr]\bigr\}^2,
\end{equation}
where $v^2=v\cdot v'$ when $v$ is an $m\times1$ vector. To evaluate
(\ref{eqnincompleteDataOIM}), we first impute $M$ samples $\bolds
\Gamma^{(i)},i=1,2,\ldots,M$, for the missing data $\bolds\Gamma$
conditional on the observed data and $\Theta$. Let $\mathbf D^{(i)} =
\bolds\Omega\cup\bolds\Gamma^{(i)}$. Then, the incomplete data
information matrix can be approximated by [\citet{989}]
%
%
\begin{eqnarray}
\label{eqnapproximateOIM} \hat{\mathcal I}(\Theta) & \doteq& \frac{1}{M}\sum
_{i=1}^{M} B\bigl(\Theta,\mathbf D^{(i)}
\bigr) - \frac{1}{M}\sum_{i=1}^M
\bigl[S\bigl(\Theta,\mathbf D^{(i)}\bigr)\bigr]^2
\nonumber\\[-8pt]\\[-8pt]\nonumber
&&{} + \Biggl[ \frac{1}{M} \sum_{i=1}^M
S\bigl(\Theta,\mathbf D^{(i)}\bigr) \Biggr]^2.
\end{eqnarray}
The SEM estimate, $\hat\Theta$, is plugged into (\ref
{eqnapproximateOIM}) to obtain $\hat{\mathcal I}(\hat\Theta)$,
which is then used to obtain the asymptotic variances of $\hat\Theta$
as well as the confidence intervals.
To ensure the accuracy of the simulation approximation, the number of
samples $M$ should be carefully chosen. The magnitude depends on the
missing data rate.

\section{Some further considerations}\label{sec4}\label{secmoreconsideration}\label{secconclusion}
Usually, products are manufactured and shipped to retailers
intermittently, meaning that the shipment dates for distinct units may
differ. Under this circumstance, we can still observe $(X, T)$ for a
returned unit. For an unreturned unit, we can subtract the date of
shipment from the end-of-study date to obtain the censored time for
$X+T$. Then, the framework discussed in Section~\ref{sec3} applies.

In some situations, direct sale from the manufacturer is possible. The
sales dates for units sold directly to customers are available in the
database. The data do not have sales lag and the lifetimes are simply
right censored. The contribution of an observed unit to the likelihood
is exactly the PDF of $T$, while if a unit is censored, say, at time
$\mathcal T_c$, the missing value can be easily imputed in the S-step as
$ t=F_T^{-1} (u+(1-u)F_T(\mathcal T_c|\Theta^{(k)}) )$,
where $F_T^{-1}(\cdot)$ is the quantile function of $F_T(\cdot)$,
while $u$ is a random draw from the uniform distribution on $(0, 1)$.

%
\begin{table}[t]
\tabcolsep=0pt
\caption{Estimated biases and RMSEs of the SEM estimator when $(\ln X,
\ln T)\sim\mathcal N(\bmu,\bSigma)$ with the consideration of a
warranty period ($\tau$) and a batch size $N = 200$}\label{tabbi-logN}
\begin{tabular*}{\tablewidth}{@{\extracolsep{\fill}}@{}lcccccccccc@{}} \hline
& & & \multicolumn{2}{c}{\textbf{True values}}\\[-6pt]
& & & \multicolumn{2}{c}{\hrulefill}\\
\textbf{Scenario} & $\bolds{\tau}$ & $\bolds{\mathcal T_{0}}$ &
\multicolumn{1}{c}{$\bmu$} & \multicolumn{1}{c}{$\bSigma$} & &
\multicolumn{1}{c}{$\bolds{\mu_1}$} & \multicolumn{1}{c}{$\bolds{\mu_2}$} &
\multicolumn{1}{c}{$\bolds{\sigma_{11}}$} & \multicolumn{1}{c}{$\bolds{\sigma_{22}}$} &
\multicolumn{1}{c}{$\bolds{\sigma_{12}}$}\\
\hline
S1 & 4 & 6 & $\pmatrix{1 \cr1}$ & $\pmatrix{1 & 0.3 \cr0.3 & 1}$ & $\matrix{ {\mbox{Bias (${\times}10^{2}$)}} \cr{\mbox{RMSE (${\times} 10$)}} }$ &
$\matrix{-0.22 \cr \phantom{-}1.68}$ &
$\matrix{-0.21 \cr \phantom{-}1.61}$ &
$\matrix{ \phantom{0}1.51 \cr \phantom{0}2.27}$ &
$\matrix{ \phantom{0}1.19 \cr\phantom{0}2.19}$ &
$\matrix{\hspace*{4pt}-2.57 \cr \phantom{-0}1.47}$
\\[8pt]
S2 & 3 & 4 & $\pmatrix{1 \cr1}$ & $\pmatrix{1 & 0.3 \cr0.3 & 1}$ &
$\matrix{ {\mbox{Bias (${\times}10^{2}$)}} \cr{\mbox{RMSE (${\times}10$)}} }$ &
$\matrix{ -2.97 \cr \phantom{-}3.16 }$&
$\matrix{ -4.40 \cr \phantom{-}2.96 }$&
$\matrix{ 15.48 \cr \phantom{0}3.82 }$&
$\matrix{ 10.33 \cr \phantom{0}3.55 }$&
$\matrix{-17.94 \cr   \phantom{-0}3.11 }$
\\[8pt]
S3 & 3 & 4 & $\pmatrix{1 \cr1.3}$ & $\pmatrix{1 & 0.4 \cr0.4 & 1}$ &
$\matrix{ {\mbox{Bias (${\times}10^{2}$)}} \cr{\mbox{RMSE (${\times}10$)}} }$ &
$\matrix{-0.50 \cr \phantom{-}3.98 }$&
$\matrix{-3.03 \cr \phantom{-}3.34 }$&
$\matrix{  \phantom{0}6.13 \cr \phantom{0}4.15 }$&
$\matrix{  \phantom{0}1.36 \cr \phantom{0}4.10 }$&
$\matrix{\hspace*{4pt}-6.24 \cr  \phantom{-0}3.15 }$\\
\hline
\end{tabular*}
\end{table}

\citet{1011} considered a nonnegligible report delay after
failure (denoted as $Y$) in addition to the sales lag $X$. When
information about~$Y$ for a returned unit is available, we can work on
the random vector $(X, Y, T)$. In the S-step, the missing $(X, Y, T)$
can be imputed similar to the acceptance-rejection method discussed in
Section~\ref{secSEMimplementation}, after which the pseudo $Q$-function can be easily specified. The M-step can be implemented
based on standard estimation procedures established for complete
multivariate data. Analysis of such data will be demonstrated in
Section~\ref{secwilsonExample}.

\section{Simulation study}\label{sec5}\label{secsimulation}
In the simulation the number of units in a batch is assumed to be $N =
200$. Both dependent and independent $(X,T)$ are examined. We first
assume a bivariate lognormal distribution for $(X,T)$:
\[
(\ln X, \ln T)\sim\mathcal N \left(\bmu= \pmatrix{\mu_{1}
\cr
\mu
_{2}}, \bSigma= \pmatrix{\sigma_{11} &
\sigma_{12}
\cr
\sigma _{12} & \sigma_{22}} \right).
\]
The biases and root mean square errors (RMSEs) of the SEM estimators
under different parameter values and different combinations of $(\tau,\mathcal T_0)$ are estimated using 5000~MC replications, as shown in
Table~\ref{tabbi-logN}.
We then consider independent $T$~and~$X$, each conforming to either an
exponential distribution or a Weibull distribution. Different settings
have been examined. The estimated biases and \mbox{RMSEs} are presented in
Table~\ref{tab2}. Code in Matlab\tsup{\textregistered} is presented in the
supplementary materials [\citet{supp}].
From Tables~\ref{tabbi-logN} and \ref{tab2}, we can see that the SEM algorithm effectively
estimates the model parameters in both dependent and \mbox{independent} cases.
We can also observe that, on average, a longer warranty period leads to
higher accuracy of the estimator. This observation agrees with our
intuition as the missing data rate decreases with $\tau$.

%
\begin{table}
\tabcolsep=0pt
\caption{Estimated biases and RMSEs of the parameter estimates
obtained from the SEM algorithm when $(X, T)$ are independent with the
consideration of a warranty period ($\tau$)}\label{tab2}
\begin{tabular*}{\tablewidth}{@{\extracolsep{\fill}}@{}lccd{2.2}cd{3.2}d{2.2}d{2.2}c@{}}
\hline
\multicolumn{3}{c}{\textbf{Setting}} & \multicolumn{2}{c}{$\bolds{\lambda}$} & \multicolumn{2}{c}{$\bolds{\theta}$} & \multicolumn{2}{c@{}}{$\bolds{\beta}$}\\[-6pt]
\multicolumn{3}{c}{\hrulefill} & \multicolumn{2}{c}{\hrulefill} & \multicolumn{2}{c}{\hrulefill} & \multicolumn{2}{c@{}}{\hrulefill}
\\
& & &
\multicolumn{1}{c}{\textbf{Bias}} & \multicolumn{1}{c}{\textbf{RMSE}} &
\multicolumn{1}{c}{\textbf{Bias}} & \multicolumn{1}{c}{\textbf{RMSE}} &
\multicolumn{1}{c}{\textbf{Bias}} & \multicolumn{1}{c@{}}{\textbf{RMSE}}
\\
$\bolds{\tau}$ & $\bolds{\mathcal T_{0}}$ & $\bolds{(\lambda, \theta, \beta)}$ &
\multicolumn{1}{c}{\textbf{($\bolds{{\times}10^{2}}$)}} & \multicolumn{1}{c}{\textbf{($\bolds{{\times}10}$)}} &
\multicolumn{1}{c}{\textbf{($\bolds{{\times}10^{2}}$)}} & \multicolumn{1}{c}{\textbf{($\bolds{{\times}10}$)}} &
\multicolumn{1}{c}{\textbf{($\bolds{{\times}10^{2}}$)}} & \multicolumn{1}{c@{}}{\textbf{($\bolds{{\times}10}$)}}
\\
\hline
\multicolumn{9}{@{}c@{}}{$X \sim \operatorname{Exp}(\lambda$), $T \sim \operatorname{Weibull}(\theta, \beta$)}\\
5 & 6 & (0.7, 5, 2) & -0.02 & 1.17 & -4.90 & 3.52 & 5.82 & 2.03 \\
3 & 4 & (0.7, 5, 2) & -5.79 & 2.71 & -42.62 & 10.18 & 20.45 & 4.45 \\
4 & 6 & (0.7, 5, 2) & -0.06 & 1.25 & -5.74 & 3.81 & 6.60 & 2.29
\\[3pt]
\multicolumn{9}{@{}c@{}}{$X \sim \operatorname{Weibull}(\theta, \beta$), $T \sim \operatorname{Exp}(\lambda$)} \\
5 & 6 & (0.5, 4, 1.5) & -0.03 & 0.77 & -4.78 & 3.39 & 4.76 & 1.54 \\
3 & 4 & (0.5, 4, 1.5) & -0.89 & 1.46 & -20.12 & 7.16 & 10.80 & 2.26 \\
4 & 6 & (0.5, 4, 1.5) & 0.02 & 0.81 & -4.11 & 3.84 & 4.28 & 1.53
\\
\hline
\end{tabular*}\vspace*{6pt}
\begin{tabular*}{\tablewidth}{@{\extracolsep{\fill}}@{}lcccccc@{}}
\multicolumn{3}{c}{\textbf{Setting}} & \multicolumn{2}{c}{$\bolds{\lambda}$} & \multicolumn{2}{c@{}}{$\bolds{\delta}$}\\[-6pt]
\multicolumn{3}{c}{\hrulefill} & \multicolumn{2}{c}{\hrulefill} & \multicolumn{2}{c@{}}{\hrulefill}
\\
$\bolds{\tau}$ & $\bolds{\mathcal T_{0}}$ & $\bolds{(\lambda, \delta)}$ &
\multicolumn{1}{c}{\textbf{Bias ($\bolds{{\times}10^{2}}$)}} & \multicolumn{1}{c}{\textbf{RMSE ($\bolds{{\times}10}$)}} &
\multicolumn{1}{c}{\textbf{Bias ($\bolds{{\times}10^{2}}$)}} & \multicolumn{1}{c@{}}{\textbf{RMSE ($\bolds{{\times}10}$)}}
\\
\hline
\multicolumn{7}{@{}c@{}}{$X \sim \operatorname{Exp}(\lambda$), $T \sim \operatorname{Exp}(\delta$)}\\
5 & 6 & (0.2, 0.2) & 1.47 & 0.63 & 1.57 & 0.63\\
4 & 5 & (0.2, 0.2) & 2.40 & 0.87 & 2.61 & 0.87\\
5 & 6 & (0.5, 0.2) & 0.74 & 0.82 & 0.50 & 0.26\\
3 & 4 & (0.5, 0.2) & 1.31 & 1.41 & 1.91 & 0.57\\
5 & 6 & (0.4, 0.7) & 0.41 & 0.37 & 0.68 & 0.72\\
3 & 4 & (0.4, 0.7) & 1.11 & 0.57 & 1.04 & 1.07\\
\hline
\end{tabular*}
\end{table}

%
\begin{table}
\tabcolsep=0pt
\caption{The estimated biases and RMSEs of Ghosh's estimators
[\citet{1014}] and the SEM estimators: $(X,T)$ are independent
and $\tau=\infty$}\label{tablepropimput}
\begin{tabular*}{\tablewidth}{@{\extracolsep{\fill}}@{}lcd{3.2}d{2.2}d{2.2}c@{}}
\hline
\textbf{Setting} & & \multicolumn{2}{c}{\textbf{Bias ($\bolds{\times}$10\tsup{2})}} &\multicolumn{2}{c@{}}{\textbf{RMSE} $\bolds{({\times}10)}$} \\[-6pt]
& & \multicolumn{2}{c}{\hrulefill} &\multicolumn{2}{c@{}}{\hrulefill}\\
$\bolds{(\mathcal T_{0}, \lambda, \theta, \beta)}$ & &
\multicolumn{1}{c}{\textbf{Impute}} &
\multicolumn{1}{c}{\textbf{SEM}} &
\multicolumn{1}{c}{\textbf{Impute}}&
\multicolumn{1}{c@{}}{\textbf{SEM}}
\\
\hline
\multicolumn{6}{@{}c@{}}{$X \sim \operatorname{Exp}(\lambda$), $T \sim \operatorname{Weibull}(\theta, \beta$)}\\
(6, 0.7, 5, 2) & $\lambda$ & -4.54 & -0.22 & 1.13 & 1.09\\
& $\theta$ & -14.95 & -1.35 & 3.83 & 3.50\\
& $\beta$ & 8.46 & 3.30 & 2.16 & 1.97
\\[3pt]
(4, 0.7, 5, 2) & $\lambda$ & -16.65 & -2.29 & 2.54 & 2.33\\
& $\theta$ & -69.54 & -9.81 & 10.21 & 8.64\\
& $\beta$ & 21.61 & 8.75 & 3.73 & 3.32
\\[6pt]
\multicolumn{6}{@{}c@{}}{$X \sim \operatorname{Weibull}(\theta, \beta$), $T \sim\operatorname{Exp}(\lambda$)}\\
(6, 0.5, 4, 1.5) & $\lambda$ & -2.93 & -0.01 & 0.60 & 0.75 \\
& $\theta$ & -21.39 & -0.75 & 2.87 & 3.80 \\
& $\beta$ & 2.47 & 2.59 & 1.36 & 1.44
\\[3pt]
(4, 0.5, 4, 1.5) & $\lambda$ & -8.53 & -0.94 & 3.13 & 1.50 \\
& $\theta$ & -47.19 & -3.41 & 26.86 & 8.33 \\
& $\beta$ & 5.31 & 6.62 & 3.91 & 2.29
\\
\hline
\\[-5pt]
\textbf{Setting} & & \multicolumn{2}{c}{\textbf{Bias $\bolds{({\times}10^{2})}$}} &\multicolumn{2}{c@{}}{\textbf{RMSE} $\bolds{({\times}10)}$} \\[-6pt]
& & \multicolumn{2}{c}{\hrulefill} &\multicolumn{2}{c@{}}{\hrulefill}\\
$\bolds{(\mathcal T_{0}, \lambda, \theta, \delta)}$ & &
\multicolumn{1}{c}{\textbf{Impute}} &
\multicolumn{1}{c}{\textbf{SEM}} &
\multicolumn{1}{c}{\textbf{Impute}}&
\multicolumn{1}{c@{}}{\textbf{SEM}}
\\
\hline
\multicolumn{6}{@{}c@{}}{$X \sim \operatorname{Exp}(\lambda$), $T \sim\operatorname{Exp}(\delta$)} \\
(5, 0.2, 0.2) & $\lambda$ & -3.45 & 0.58 & 1.59 & 0.70\\
& $\delta$ & -2.09 & 1.83 & 0.94 & 0.76
\\[3pt]
(4, 0.4, 0.7) & $\lambda$ & -3.52 & 0.44 & 1.36 & 0.51\\
& $\delta$ & -2.64 & 0.42 & 1.19 & 1.01\\
\hline
\end{tabular*}
\end{table}

The proportional imputation method proposed by \citet{1014} does
not allow for a warranty limit and it can only handle independent $X$
and $T$. In order to compare the SEM algorithm with it, we let $X$ and
$T$ be independent and $\tau> \mathcal T_{0}$ (i.e., no warranty
consideration). The biases and RMSEs of the estimators computed from
the proportional imputation approach and the SEM algorithm are
presented in Table~\ref{tablepropimput}. The SEM estimator has much
smaller biases and RMSEs. A possible explanation is that the stratified
sampling scheme in the proportional imputation algorithm might
introduce biases in the imputing samples. Another finding from our
comparative study is that the computation time required by the SEM
algorithm is much shorter compared to that of the proportional
imputation algorithm. Overall, the SEM algorithm is statistically and
computationally more efficient than the imputation method. More
importantly, the SEM algorithm is able to handle a more general
scenario with a warranty limit and dependent $X$ and $T$. It also
allows for construction of confidence intervals for the parameters.
These advantages make the SEM algorithm attractive for the problem.

%
\begin{figure}

\includegraphics{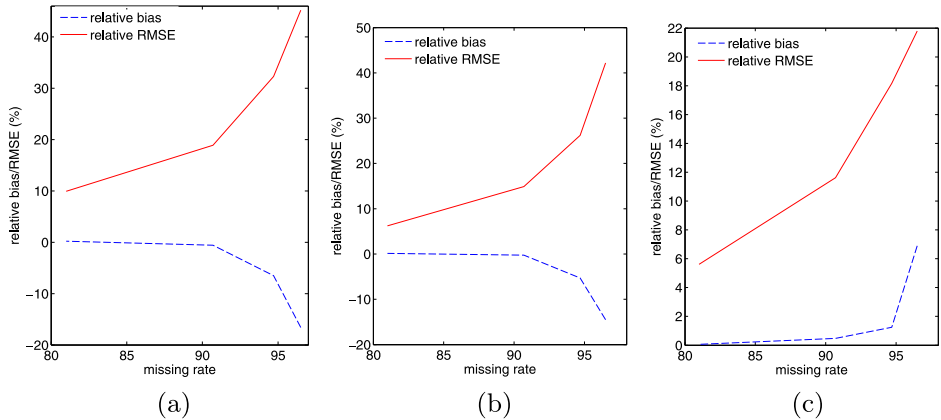}

\caption{Relative biases and relative RMSEs of the SEM estimates under
the Exp--Weibull setting.
\textup{(a)}~Is for $\lambda$,
\textup{(b)}~is for $\theta$, and
\textup{(c)} is for $\beta$.} \label{fighighcensor}
\end{figure}

%
\begin{figure}[b]

\includegraphics{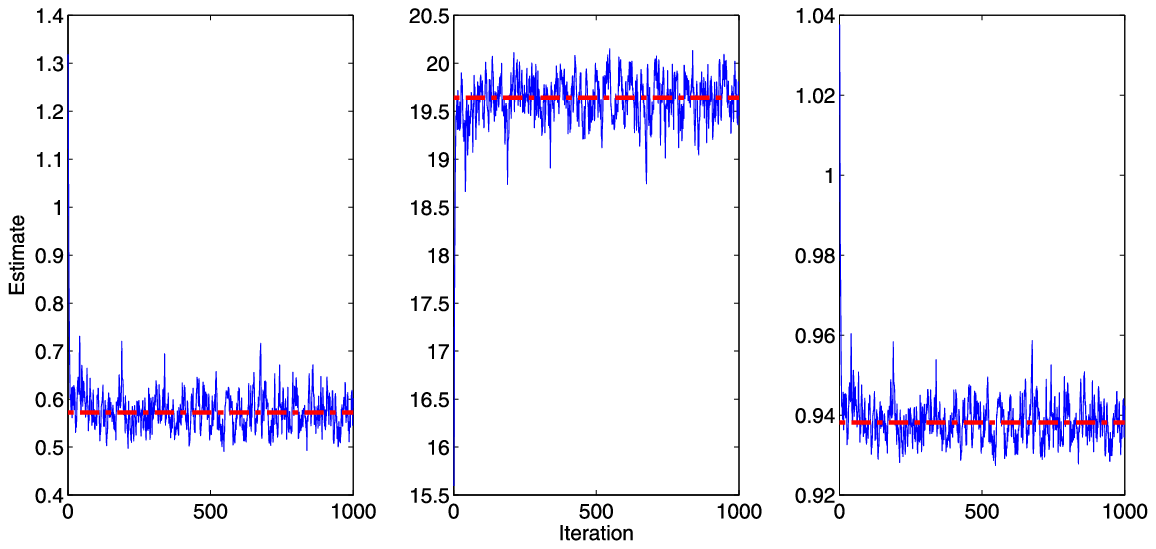}

\caption{Parameter evolutions in the SEM algorithm when there is no
warranty: the dashed-dotted line represents the average of the last 900
iterations.}\vspace*{-5pt}\label{figfurnaceSEM}
\end{figure}

To demonstrate the advantage of the
SEM algorithm over direct optimization, further simulation is conducted
by assuming $X \sim\operatorname{Exp}(\lambda=0.7)$, $T \sim
\operatorname{Weibull}(\theta,\beta=2)$, and $N = 2000$. Different
missing data rates are achieved by varying $\theta$. We find that
direct maximization breaks down very quickly (i.e., fails to converge)
when the missing data rate is high, say, $>$80\%. On the other hand,
the SEM algorithm performs well under much larger missing data rates.
The relative biases (bias $\div$ true value) and relative RMSEs (RMSE
$\div$ true value) of the SEM estimators are computed from 1000~MC
replications, as shown in Figure~\ref{fighighcensor}. When the missing
data rate is extremely high, say, 97\% in Figure~\ref
{fighighcensor}, the RMSE for~$\theta$ is large, which can be seen as
a breakdown of the SEM algorithm. For a fixed missing data rate,
nevertheless, the bias and RMSE can be significantly reduced if the
sample size $N$ is increased. For illustration, given the missing data
rate of 94.7\%, the respective relative biases (RMSEs) for $\lambda,\theta$ and $\beta$ decrease from $-$6.5\% (32.2\%), $-$5.3\% (26.1\%),
1.25\% (18.2\%) to $-$2.0\% (15.1\%), $-$1.8\% (13.2)\%, 0.15\% (6.0\%), respectively, when $N$ is increased to 20,000.

\section{Examples}\label{sec6}\label{secexample}
The developed algorithm is applied to three real data sets with
different missing data patterns. The first example is from an
industrial firm that produces residential furnace components
[\citet{1014}]. There is an unobserved sales lag for an
unreturned unit but there is no warranty limit. The second example
comes from warranty claims for an automobile component with both a
sales lag and a warranty limit. The third example concerns a
telecommunications product [\citet{1011}] where both the sales
lag and report delay exist. The times in these examples are in months.
These data are presented in the supplementary materials [\citet{supp}].

\subsection{Installation failure data of a furnace}\label{sec6.1}

This data set is from an industrial firm producing residential furnace
components during one week in May 2001. It consists of $N=400$ furnace
components and $C = 133$ returns, denoted as $(x_i, t_i)$ for $i =
1,\ldots, C$. The components are sold with life warranty, that is,
$\tau=\infty$. In keeping with \citet{1014}, suppose the sales
lag is exponential, $X \sim \operatorname{Exp}(\lambda$), and the failure time is
Weibull, $T \sim \operatorname{Weibull}(\theta, \beta$). \citet{1014}
obtained estimates of the model parameters as $\hat\lambda = 0.57$,
$\hat\theta =14.47$, and $\hat\beta= 0.81$ by using his imputation
algorithm. Here, we reanalyze the data using the SEM algorithm. We use
100 iterations for burn-in and another 900 iterations to obtain the SEM
estimates. The evolution paths of the parameters are shown in
Figure~\ref{figfurnaceSEM}. The paths reveal no obvious trend in the
simulation. The computation time for the SEM algorithm is 12.28 seconds
on a laptop with an Intel\tsup{\textregistered} Core i5 CPU, which is
faster than that required by the proportional imputation method (72.12
seconds on the same computer). We then invoke the procedure in
Section~\ref{secCIconstruction} to compute the information matrix and
thus the standard deviations of the estimators. To ensure an accurate
approximation for the information matrix, we use $M ={}$100,000
imputations in (\ref{eqnapproximateOIM}).
The estimates\vspace*{1pt} (standard errors) of the model parameters are $\hat
\lambda= 0.57$ (0.053), $\hat\theta= 19.59$~(2.420), and $\hat\beta
= 0.95$ (0.078), respectively.

\subsection{Warranty data for an automobile component}\label{sec6.2}

The data analyzed here are warranty claims for a specific automobile
component produced over a three-year period. The component is sold with
an 18-month warranty. When a component fails within warranty and is
returned as a claim, the date of manufacture, date of sale, date of
claim, failure mode, and some other related information are recorded.
The end-of-study date for this study is $\mathcal T_0 = 54$ months. We
focus on the 589 components manufactured in the first month (month 0)
of the production. During the observation window, 66 claims were observed.

\begin{figure}[t]

\includegraphics{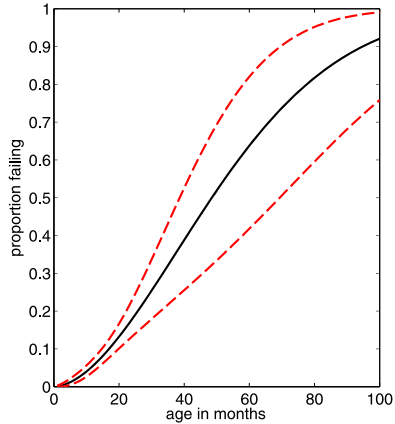}

\caption{Estimated CDF and 95\% pointwise confidence band for the
failure time $T$ for the automobile component.}\label{figCDFautoComp}
\end{figure}

Based on previous experience, we use a lognormal distribution for the
sales lag and Weibull for the lifetime, that is, $X \sim\ln\mathcal
N(\mu, \sigma)$ and $T\sim\operatorname{Weibull}(\theta, \beta)$.
To ensure convergence of the SEM algorithm, 100,000 iterations are
used. The running time is about 10 minutes. The evolution paths of the
parameter estimates versus the SEM iterations are presented in the
supplementary materials [\citet{supp}].
The estimates (standard errors) of the four model parameters are $\hat
\mu= 1.66$~(0.107), $\hat\sigma= 0.84$ (0.081), $\hat\theta= 59.5$
(10.0), and $\hat\beta= 1.79$ (0.224), respectively. The estimated
lifetime distribution and the corresponding 95\% pointwise confidence
band are depicted in Figure~\ref{figCDFautoComp}. One can also obtain
estimates of reliability characteristics [e.g., mean time to failure
(MTTF), quantiles, etc.], which are useful in improving product
reliability as well as determining the optimal warranty period.

To check the parametric model assumption, we consider different
combinations of the distributions for $(X,T)$. The log-likelihood at
the estimated values of the model parameters and the AIC are presented
in Table~\ref{tabmodelcomp}. A lognormal distribution for the sales
lag and a Weibull distribution for the component lifetime seems reasonable.

\subsection{Field data for a telecommunications product}\label{sec6.3}\label{secwilsonExample}

\citet{1011} reported field failure data for a product installed
in a telecommunications network. The data consist of 1838 units in
total, out of which 26 units were returned within $\mathcal T_0 = 18$
months after the shipment. The failure data are grouped by month so
that we only observe the number of failures for each month. All the
remaining 1812 units are missing and the missing proportion is about
98.6\%. In this data set the recorded time for each of the 26 returned
units is the time in between the unit being shipped and being returned
for repair.
The recorded time includes the sales lag $X$, failure time $T$, and
report delay {$Y$}. This means that a failure is recorded only when $X
+ T + Y < {\mathcal T}_0$. But if a failure is recorded, we only
observe {$X+T+Y$}. In order to decouple these three random variables,
\citet{1011} collected additional sales-lag data and report-delay
data from an old product in the same family, for which the sales lag
and the return delay are assumed to be the same as the product under
study. In total, there are 100 extra installation-lag data and 100
extra report-delay data records. These two data sets are also interval
censored and grouped by month. More details about the data can be found
in the original paper.

%
%

%
\begin{table}[b]
\tabcolsep=0pt
\caption{Values of the likelihood and Akaike's information criterion
(AIC) under different parametric models in Example 6.2}\label{tabmodelcomp}
\begin{tabular*}{\tablewidth}{@{\extracolsep{\fill}}ld{4.1}d{4.1}d{4.1}d{4.1}@{}}
\hline
& \multicolumn{1}{c}{$\bolds{X \sim \operatorname{Exp},}$}
& \multicolumn{1}{c}{$\bolds{X \sim \operatorname{Weibull},}$}
& \multicolumn{1}{c}{$\bolds{(X, T)\sim }$}
& \multicolumn{1}{c@{}}{$\bolds{X \sim \operatorname{Logn},}$}
\\
\textbf{Model}
& \multicolumn{1}{c}{$\bolds{T \sim \operatorname{Weibull}}$}
& \multicolumn{1}{c}{$\bolds{T \sim \operatorname{Weibull}}$}
& \multicolumn{1}{c}{$\bolds{\operatorname{Bivariate} \operatorname{Logn}}$}
& \multicolumn{1}{c@{}}{$\bolds{T \sim \operatorname{Weibull}}$}\\
\hline
No. of parameter & 3 & 4 & 5 & 4 \\
Likelihood & -586.5 & -584.2 & -578.1 & -578.5 \\
AIC & 1179.0 & 1176.4 & 1166.2 & 1165.0 \\
\hline
\end{tabular*}
\end{table}

\citet{1011} pointed out that direct maximum likelihood
estimation is difficult. They developed a Bayesian inference procedure
to fit the data. The Gibbs sampling was adopted to resemble the
posterior distribution.
Here, we apply the SEM algorithm.
Following \citet{1011}, we assume $X \sim\operatorname
{Gamma}(k_1, \lambda_1)$, where $k_1$ is the shape parameter and
$\lambda_1$ is the scale parameter. The failure time is assumed to be
Weibull, that is, $T \sim\operatorname{Weibull}(\theta, \beta)$,
and the report delay is gamma, that is, $ {Y} \sim\operatorname
{Gamma}(k_2, \lambda_2)$. We fit the additional 100 installation-lag
data and the additional 100 report-delay data to obtain an initial
estimate of $k_1, \lambda_1$ and $k_2, \lambda_2$. These values are
used as initial values for the SEM algorithm. In the SEM iterations, we
impute the missing $X$, $T$, and $Y$ based on the fact that $X$ and $Y$
in the additional data sets and $X + Y + T$ in the original data set
are interval censored or right censored. The imputation can be done by
the acceptance-rejection method with acceptance only when the imputed
value falls inside the desired interval.
Since the missing proportion is high, we use 100,000 iterations in the
SEM algorithm. The first 10,000 iterations are discarded for burn-in
purposes and the remaining 90,000 iterations are averaged for
estimation. The parameter estimates are $\hat k_1 = 2.264$ (0.40),
$\hat\lambda_1 = 1.714$ (0.35), $\hat\theta= 720.7$\vspace*{2pt} (683), $\hat
\beta= 1.153$ (0.37), $\hat k_2 = 2.779$ (0.49), and $\hat\lambda_2
= 0.080$ (0.015).
The evolution paths of the six parameters are presented in the
supplementary materials [\citet{supp}].
The traces for the parameters related to the lifetime $T$ are very
unstable. This can be viewed as an indication of large bias/variance in
the estimation or an indicator of insufficient information for $T$,
which might lead to the breakdown of SEM. With these results, one can
decide whether a longer observation window is needed. In summary, the
SEM algorithm serves well as a tool for checking whether there is
sufficient information for inference.


%

\section{Conclusions}\label{sec7}\label{secconclusion}
The common problem of unknown sales dates in field failure data has
posed a challenge. Direct maximization of the likelihood is difficult
due to the excessive flatness of the likelihood and numerical error
when evaluating the function. We have proposed an SEM framework for
parametric inference. The algorithm allows for a warranty limit and
possible dependence between the sales lag and the product lifetime. It
is easy to implement and computationally efficient. Our examples with
different missing data patterns demonstrate the flexibility of the
proposed framework.

\section*{Acknowledgments}
The authors are grateful to the Editor and three anonymous reviewers
for constructive comments which have led to a substantial \mbox{improvement}
to an earlier version of the paper. They would like to thank Dr.~S.~Wilson and Dr. T. Joyce for sharing the telecommunications product
data. They are also grateful to Dr. R. Karim for the discussion of
sales-lag problems in warranty data analysis and Dr. W. A. Woodward for
comments on the paper.

\begin{supplement}
\stitle{Additional discussions, graphs, Matlab code, and data\\}
\slink[doi]{10.1214/14-AOAS752SUPP} 
\sdatatype{.pdf}
\sfilename{aoas752\_supp.pdf}
\sdescription{We provide additional discussions on the effect of model
misspecification and evolution paths of parameter estimates in SEM for
Sections~\ref{sec6.2} and \ref{sec6.3}. We also provide the Matlab code for simulation
and the data used in the examples.}
\end{supplement}


\printaddresses
\end{document}